\documentclass[prb,footinbib,showpacs,twocolumn,amsmath,amssymb]{revtex4}
\usepackage{graphicx}% 
\usepackage{xcolor}
\usepackage{bm}
\usepackage{amsmath, amsthm, amssymb}
\usepackage{braket}
\renewcommand\Im{\operatorname{Im}}
\renewcommand\Re{\operatorname{Re}}

\usepackage[normalem]{ulem}
\usepackage{upgreek}

\newcommand{\bek}{\begin{eqnarray}}
\newcommand{\ek}{\end{eqnarray}}
\newcommand{\VEC}[1]{\mathbf{#1}}
\newcommand{\ud}{\mathrm{d}}
\newcommand{\iu}{\mathrm{i}}
\usepackage[caption=false,margin=0pt]{subfig}

\usepackage{array,booktabs}
\usepackage[version=4]{mhchem}

\definecolor{orange}{rgb}{1,0.5,0}
\usepackage{scrextend}

\begin{document}

\title{Longitudinal and transverse spin relaxation times of magnetic single adatoms: \\
an \textit{ab initio} analysis}

\author{Julen Iba\~{n}ez-Azpiroz, Manuel dos Santos Dias, Stefan Bl\"ugel, Samir Lounis}
\address{Peter Gr\"unberg Institute and Institute for Advanced Simulation, Forschungszentrum
J\"ulich \& JARA, D-52425 J\"ulich, Germany}
\date{\today}

%\pacs{}

\begin{abstract}
We present a systematic \textit{ab initio} investigation of the longitudinal 
and transverse spin relaxation times of magnetic single adatoms deposited on metallic substrates. 
Our analysis based on time-dependent density
functional theory shows that the longitudinal time, $T_{\parallel}$, is of order femtosecond 
while the transverse time, $T_{\perp}$, is of order picosecond, i.e.
$T_{\perp}\gg T_{\parallel}$.
This comes as a consequence of the different energy scales of the corresponding processes:
$T_{\parallel}$ involves spin-density excitations of order eV, while
$T_{\perp}$ is governed by atomic spin-excitations of order meV.
Comparison  to  available  
inelastic scanning tunneling spectroscopy $\mathrm{d}I/\mathrm{d}V$ 
experimental curves  shows that
the order of magnitude of $T_{\perp}$ agrees well with our results. 
Regarding $T_{\parallel}$, 
the time scale calculated here is 
several orders of magnitude faster than what has been measured up to now;
we therefore propose that 
an ultrafast laser pulse measuring technique is required in order
to access the ultrafast spin-dynamics described in this work.
\end{abstract}
\maketitle

\section{Introduction}

Single adatoms deposited on substrates offer an exceptional 
scenario for studying magnetism at the atomic scale, given that these tiny objects 
can develop a large magnetic moment of 
several Bohr magnetons~\cite{oswald_giant_1986,wildberger_magnetic_1995,lang_local_1994}
as well as a large magnetic anisotropy energy 
barrier of few meV~\cite{gambardella_giant_2003,rau_reaching_2014,PhysRevLett.108.256811,
heinrich_single-atom_2004,hirjibehedin_spin_2006,
heinrich_tuning_2015,PhysRevLett.111.157204,PhysRevLett.114.106807,oberg_control_2014},
both extremely desirable properties for potential applications in spintronic devices.
Interestingly, the possibility of tuning and engineering these and 
other properties by the suitable combination 
of adatom and substrate material (possibly including coating layers) provides plenty of room
for research in this area.

In order to achieve the ultimate goal of a technologically applicable magnetic single adatom, however, 
not only the static properties need to be adequate but also 
the dynamical ones, and in particular the ones related to the spin, i.e. the 
spin-dynamics. 
For example, fast spin-dynamics can  be useful when the goal is to transfer magnetic information
from or to the adatom, while slow spin-dynamics are desirable if 
the aim is to store magnetic information.
In comparison to the static case, the study of spin-dynamics of single adatoms 
is much more recent and has only hatched out after the 
advent of   
spin-polarized scanning tunneling microscopy (STM)
and  
inelastic electron tunneling spectroscopy (IETS).
These experimental techniques, occasionally used in combination with X-ray  
magnetic circular dichroism (XMCD)~\cite{donati_magnetic_2016} and 
electron paramagnetic resonance (EPR)~\cite{baumann_electron_2015},
allow  to monitor the 
dynamical regime by, \textit{e.g.}, measuring atomic 
spin-excitations~\cite{heinrich_single-atom_2004,hirjibehedin_large_2007,otte_role_2008,PhysRevLett.106.037205}
and quasiparticle interferences~\cite{PhysRevLett.107.186805},
accessing spin relaxation times~\cite{loth_measurement_2010,baumann_electron_2015}
and even resolving highly dynamical processes like the reading and writing of 
magnetic information into a 
single adatom~\cite{natterer_reading_2017}.

From the theoretical point of view, spin-dynamics of single adatoms
have also attracted a great deal of attention in the past few years. 
In this context, time-dependent 
density functional theory~\cite{PhysRevLett.52.997} 
(TDDFT) has proven to be a powerful tool for characterizing the 
spin-excitation spectrum and, more generally, 
giving insight into the connection between what is measured experimentally
and the underlying electronic structure (see, \textit{e.g.}, Refs.~\onlinecite{lounis_dynamical_2010,lounis_theory_2011,dias_relativistic_2015,PhysRevB.91.104420,PhysRevB.89.235439,ibanez-azpiroz_zero-point_2016}).
Alongside, model Hamiltonians have also been used to
analyze, among other aspects, the role of symmetry on the switching rate of the magnetic moment
~\cite{hubner_symmetry_2014,khajetoorians_current-driven_2013},
electron tunneling processes in IETS 
experiments~\cite{lorente_efficient_2009,PhysRevLett.102.256802,PhysRevLett.103.050801,fransson_spin_2009,sothmann_nonequilibrium_2010}
and spin-decoherence~\cite{delgado_spin_2010,oberg_control_2014,delgado_spin_2017}.

In this paper,
we present an \textit{ab initio} study based on density functional theory (DFT) and TDDFT 
of two relaxation processes of
single adatoms, namely the longitudinal and transverse 
spin relaxations characterized by the relaxation times $T_{\parallel}$ and $T_{\perp}$, respectively. 
Physically, 
$T_{\parallel}$ characterizes the relaxation of the size of the adatom's spin magnetic moment
while $T_{\perp}$ describes its damped precessional motion.
Employing \textit{ab initio}-derived expressions, 
we systematically provide hard numbers for $T_{\parallel}$ and $T_{\perp}$ 
for a series of 3\textit{d} and 4\textit{d} transition metal adatoms deposited on two metallic
substrates, namely Ag(100) and Cu(111). 
Our analysis shows that, while $T_{\parallel}$ is of the order of femtosecond, 
$T_{\perp}$ ranges from few to thousands picoseconds, i.e., $T_{\perp}\gg T_{\parallel}$. 
Noteworthily, these time scales are settled by the corresponding energy scales of the
associated processes; continuous spin-conserving single-particle excitations of energy eV
in the case of $T_{\parallel}$, atomic spin-flip spin-excitations of energy meV in the case
of $T_{\perp}$.
In comparison to available experimental measurements, 
the relaxation times $T_{\perp}$ extracted from IETS $\text{d}I/\text{d}V$ curves show overal
the same order of magnitude as the ones calculated in our work, 
and agree remarkably well in
specific cases such as Fe on Cu(111)~\cite{PhysRevLett.106.037205}.
Regarding $T_{\parallel}$, 
the time resolution of the 
currently available measuring techniques  
ranges from few nanoseconds to hundreds of picoseconds~\cite{loth_measurement_2010,saunus_versatile_2013}, 
hence not enough to  
monitor the femtosecond regime predicted here.
However, considering the technological developments within 
this field~\cite{kruger_attosecond_2011,cocker_ultrafast_2013,cocker_tracking_2016}, access to the fs 
time scale of magnetic adatoms could be realized in the near future, thus giving access to the spin-dynamics
described in this work.

The paper is organized as follows. Sec. \ref{sec:compu} summarizes the technical
details of the formalism used throughout the work. 
In Sec. \ref{sec:GS} we present DFT calculations of ground state properties of
several 3\textit{d} and 4\textit{d} transition metal adatoms deposited on Ag(100) and Cu(111).
In Sec. \ref{sec:lifetimes} we extend the analysis to the dynamical regime; 
in particular, we calculate 
longitudinal (Sec. \ref{subsec:long}) and transverse (Sec. \ref{subsec:trans})
relaxation times within the TDDFT framework. 
Conclusions and a summary  of the main results are provided in Sec. \ref{sec:discussion}. 
In Appendices \ref{appendix:Bloch-long} and  \ref{appendix:LLG}
we derive the connection between TDDFT and phenomenological
models for the longitudinal and transverse dynamics, respectively.
Finally, Appendix \ref{appendix:BR} contains a short summary of the Bloch-Redfield formalism
in order to allow comparison of our TDDFT-based work to 
other theoretical  analyses.
 
\begin{figure}[t]
\includegraphics[width=\columnwidth]{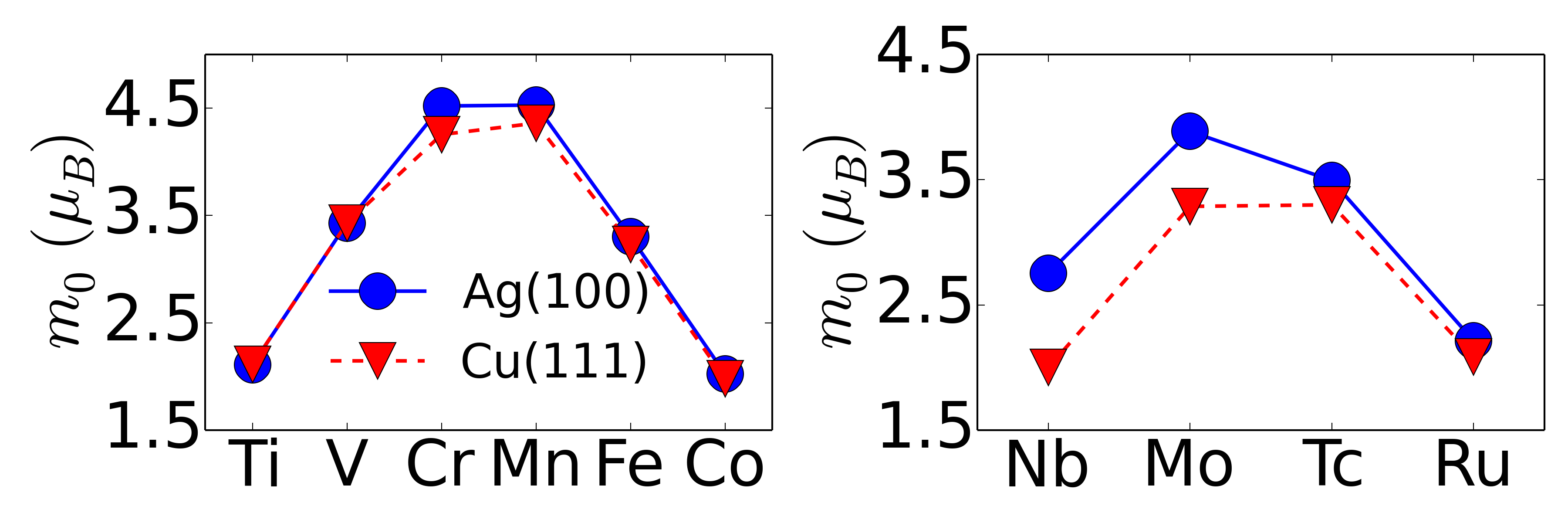}
\caption{(color online) Calculated ground state spin magnetic moments for several 3\textit{d} (left) and 4\textit{d} (right)
transition metal adatoms deposited on Ag(100)  and Cu(111) denoted by circles (blue) and
triangles (red), respectively.
}
\label{fig:M0}
\end{figure}

\section{Computational details}
\label{sec:compu}

We have performed DFT calculations using 
the Korringa-Kohn-Rostoker Green function (KKRGF) approach, employing  
the atomic sphere approximation 
with full charge density~\cite{papanikolaou_conceptual_2002}
including spin-orbit coupling~\cite{dias_relativistic_2015} (SOC). 
Exchange and correlation (XC) effects have been taken into account using the local 
spin-density approximation  with the parametrization by 
Vosko, Wilk and Nusair~\cite{vosko_accurate_1980}. 
We have modeled the two surfaces Ag(100) and Cu(111) 
using a slab composed of 24 layers and 
augmented by 
two vacuum regions of 21.1 \AA$\;$ thickness each, 
employing the lattice constants $a = 5.46$ \AA$\;$  and $a =4.83$ \AA, respectively.
The vertical distance from adatom to the surface layer has been 
calculated using the structural relaxation scheme
implemented in the QUANTUM-ESPRESSO package~\cite{espresso}, considering the 
convergence criterion whereby forces are $<10^{-4}$ Ry a.u.$^{-1}$ and employing
norm-conserving pseudopotentials, a $4\times4$ two-dimensional unit cell,
$\Gamma$ point calculation and a cutoff energy of 80 Ry.
In all cases, the distance between adatom and substrate was reduced
by approximately 15$\%$ with respect to the ideal value. Hence, for the sake of comparison, 
we adopted the same distance for all adatoms in the DFT and TDDFT calculations 
using the KKRGF method. 
Noteworthily, this method allows a real-space treatment of 
the adatoms through an embedding technique~\cite{papanikolaou_conceptual_2002}.
Following this scheme, we have employed converged real-space 
clusters of 43 and 55 sites for the Ag(100) and Cu(111) 
surfaces, respectively.

\section{Ground state properties}
\label{sec:GS}

\begin{figure}[t]
\includegraphics[width=\columnwidth]{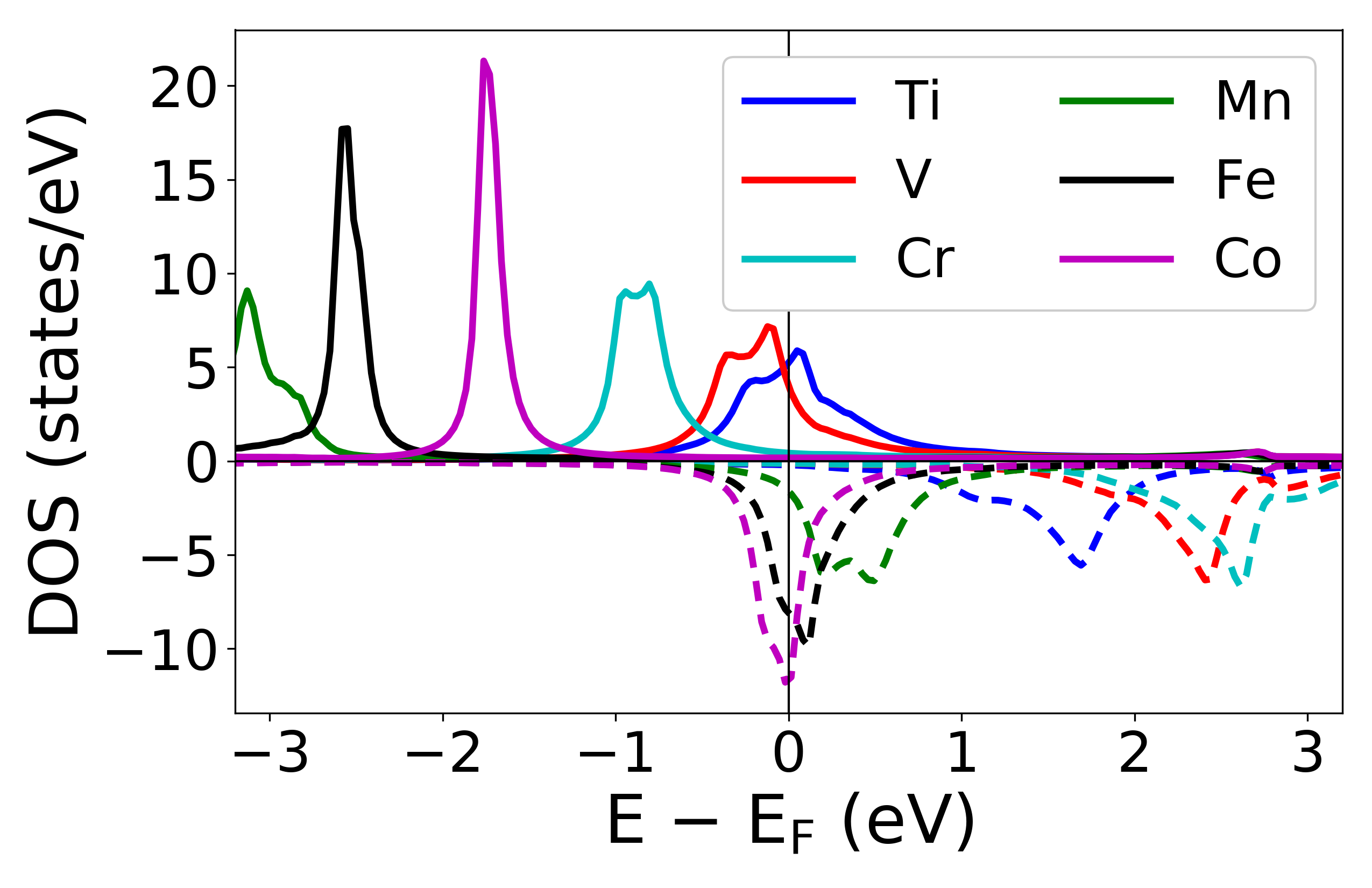}
\caption{(color online) Atom-projected total electronic DOS for 3\textit{d} adatoms deposited on Ag(100).
The majority and minority spin channels are denoted by solid (positive) and
dashed (negative) lines, respectively. The vertical line denotes the Fermi level. 
}
\label{fig:DOS}
\end{figure}

In this section we analyze two ground state properties, namely the spin 
magnetic moment,  denoted by $m_{0}$, and electronic 
density of states (DOS) 
of several 3\textit{d} and 4\textit{d} transition metal adatoms deposited on the metallic substrates Ag(100) 
and Cu(111). Let us begin with Fig. \ref{fig:M0}, where the calculated  $m_{0}$ is depicted. 
This figure shows that all the considered  adatoms develop large magnetic moments 
of more than $2$ $\mu_{B}$. Furthermore,  $m_{0}$ acquires non-integer values, 
indicating the itinerant character of the 
adatom's \textit{d} electrons
induced by the hybridization with the electrons of the metallic substrate. 
This feature 
is confirmed by the DOS, which is  displayed in Fig. \ref{fig:DOS} for the specific case of 3\textit{d} 
adatoms deposited on Ag(100).
This figure shows that the \textit{d}-state peaks, so-called virtual bound states, 
are substantially broadened 
(between $\sim0.1$ eV and $\sim1$ eV depending on the adatom),
which is a well-known consequence of 
hybridization with 
the substrate~\cite{wildberger_magnetic_1995,PhysRevLett.106.037205,dias_relativistic_2015}.
A further property indicated  by  Fig. \ref{fig:M0} is that the first of the
atomic Hund's rules is closely fulfilled, i.e.
the half filled \textit{d}-shell elements develop the largest magnetic moments, case of
Cr and Mn for 3\textit{d}, Mo and Tc for 4\textit{d}.
Finally, Fig. \ref{fig:M0} shows that the choice of metallic substrate and surface orientation
does not substantially affect the spin magnetic moment developed by the adatom, indicating that 
the symmetry of the substrate plays a minor role in this context. 
These ground state properties are consistent with the original works by 
Dederichs and co-workers~\cite{oswald_giant_1986,wildberger_magnetic_1995,lang_local_1994},
as well as with more recent studies~\cite{lounis_dynamical_2010,lounis_theory_2011,dias_relativistic_2015,PhysRevB.91.104420,PhysRevB.89.235439,ibanez-azpiroz_zero-point_2016}.

\section{Spin-susceptibilities and relaxation times}
\label{sec:lifetimes}

In this section we analyze  dynamical properties of the magnetic 
adatoms studied in the previous section,
paying special attention to  relaxation times and their connection to 
the electronic structure. For this, let us consider the linear response of a ferromagnetic
system to an externally applied time-dependent perturbation,
\bek\label{eq:general-response}
\delta {\textbf{m}}(\textbf{r};t)=
\int d\textbf{r}'\int dt'
\boldsymbol{\chi}(\textbf{r},\textbf{r}';t-t')\delta {\textbf{V}}(\textbf{r}';t').
\ek
Above, $\delta{\textbf{m}}=(\delta m_{x},\delta m_{y},\delta m_{z},\delta n)$ and 
$\delta{\textbf{V}}=(\delta B_{x},\delta B_{y},\delta B_{z},\delta V)$, with 
$\delta m_{i}$ and $\delta B_{i}$ respectively 
the components of the spin magnetic moment and external magnetic field, while
$\delta n$ and $\delta V$ are the charge density and external scalar field, respectively. 
In frequency space and defining atomic-like quantities by integrating 
out the spatial dependence over atomic sites~\cite{lounis_theory_2011}, the above expression
takes the simplified form
\bek\label{eq:general-response-w}
\delta {\textbf{m}}(\omega)=
\boldsymbol{\chi}(\omega)\delta {\textbf{V}}(\omega).
\ek
The quantity $\boldsymbol{\chi}$ in the above equations is a 4$\times$4
tensor that couples in general all components of the spin and charge responses
with each other.
If SOC is weak, however,
the full response decouples into a longitudinal and transverse
part~\cite{book_vignale}. 
This approximation is justified for the systems investigated here since the 
off-diagonal sectors of the susceptibility tensor are small in comparison to the diagonal ones.
Then, assuming that the perturbation is purely of
magnetic origin (i.e. $\delta V=0$), 
the change of the spin magnetic moment length
is described by
\bek\label{eq:dmz-w}
\delta m_{z}(\omega)=
\chi_{\parallel}(\omega)\delta B_{z}(\omega).
\ek
Above, $\chi_{\parallel}(\omega)$ 
denotes the longitudinal spin-susceptibility.
This quantity is determined by excitations between electrons with same spin state,
given that it involves the Pauli matrix $\sigma_{z}$ that is
diagonal in spin basis~\cite{PhysRevLett.119.017203}.
On the other hand, the change of the transverse spin components 
can be compactly described using the circular combinations $m_{\pm}=m_{x}\pm im_{y}$ and 
$B_{\pm}=B_{x}\pm iB_{y}$, yielding for the $+$ component
\bek\label{eq:dm+_w}
\delta m_{+}(\omega)=
\chi_{\pm}(\omega)\delta B_{+}(\omega).
\ek
Above, $\chi_{\pm}(\omega)$ denotes the transverse spin-susceptibility which, contrary to 
$\chi_{\parallel}(\omega)$, is determined by transitions that flip the spin state of the electrons
due to the transverse Pauli spin matrices involved, 
which are off-diagonal in spin space~\cite{lounis_dynamical_2010,lounis_theory_2011,dias_relativistic_2015}. 

In the following, the analysis is divided in two subsections:
Sec. \ref{subsec:long} deals with the longitudinal response while Sec. \ref{subsec:trans} 
deals with the transverse component.

\subsection{Longitudinal component}
\label{subsec:long}

The general expression for the adatom's enhanced longitudinal spin-susceptibility 
(see Eq. (\ref{eq:dmz-w})) within the
TDDFT framework~\cite{PhysRevLett.119.017203} 
is given by
\begin{equation}
\label{eq:susc-long}
\chi_{\parallel}(\omega)=\dfrac{\chi^{KS}_{\parallel}(\omega)}{1-U_{\parallel}\cdot \chi_{\parallel}^{KS}(\omega)},
\end{equation}
where $U_{\parallel}$ denotes the longitudinal XC kernel treated in the 
adiabatic local spin-density approximation~\cite{ortenzi_accounting_2012} 
including the Coulomb term, while 
$\chi^{KS}_{\parallel}(\omega)$ is the longitudinal KS spin-susceptibility.
We note that neglecting the direct contribution of the substrate atoms to the magnetic
spin-susceptibility is justified in the Ag and Cu substrates analyzed here
since the polarizability of such elements is 
very weak~\cite{lounis_dynamical_2010,lounis_theory_2011,dias_relativistic_2015,PhysRevB.91.104420,PhysRevB.89.235439,ibanez-azpiroz_zero-point_2016}.

In essence,
$\chi_{\parallel}(\omega)$ in Eq. (\ref{eq:susc-long}) describes the ability of the system
to continuously modify the size of its magnetic moment by an externally applied
time-dependent magnetic perturbation along the magnetization direction.
The dynamics of this process can be 
phenomenologically studied in terms of the longitudinal Bloch equation, 
which yields the following form for the enhanced spin-susceptibility~\cite{white_quantum_2007}
(see Appendix \ref{appendix:Bloch-long}),
\bek\label{eq:bl-Tlong}
\chi^{\text{Bl}}(\omega) =
\dfrac{\chi^{\text{Bl}}_{0}}{1-i\omega T_{\parallel}}.
\ek
Above, $\chi^{\text{Bl}}_{0}$ denotes a static spin-susceptibility, while 
$T_{\parallel}$ corresponds to the longitudinal relaxation time mentioned in the introduction. 
Our aim is to establish a direct comparison between 
Eqs. (\ref{eq:susc-long}) and (\ref{eq:bl-Tlong}). For this purpose, let us
use the first-order Taylor expansion of the KS spin-susceptibility~\cite{PhysRevLett.119.017203} 
\bek\label{eq:chi-KS-long}
\chi^{KS}_{\parallel}(\omega)\simeq \rho_{F}   
-in_{e\text{-}h}\omega,
\ek
with  $\rho_{F}=\rho_{F,\uparrow}+\rho_{F,\downarrow}$ the DOS at the Fermi level
and $n_{e\text{-}h}=\pi(\rho^{2}_{F,\uparrow}+\rho^{2}_{F,\downarrow})/2$ the density of electron-hole
excitations of the same spin channel.
By inserting $\chi^{KS}_{\parallel}(\omega)$ of Eq. (\ref{eq:chi-KS-long})
into Eq. (\ref{eq:susc-long}),
$\chi_{\parallel}(\omega)$ acquires a functional form in $\omega$
equal to that of $\chi^{\text{Bl}}(\omega)$ in Eq. (\ref{eq:bl-Tlong}). This then allows to obtain an 
expression for the longitudinal relaxation time in terms of
basic electronic properties  
(see Appendix \ref{appendix:Bloch-long} for details):
\begin{equation}
\label{eq:Tlong}
T_{\parallel}=
\dfrac{U_{\parallel}n_{e\text{-}h}}{U_{\parallel}\rho_{F}-1}.
\end{equation}
The above expression is one of the main results of the present work. 
First of all, it shows that the longitudinal relaxation time is 
settled by the magnitude of electron-hole excitations weighted by
the XC kernel (see the denominator of Eq. (\ref{eq:Tlong})),
both quantities of order eV, hence settling the time scale of 
$T_{\parallel}$ as fs.
Secondly, it shows that $T_{\parallel}$ diverges as 
$U_{\parallel}\rho_{F}\rightarrow 1$ (see the unitless denominator
in the equation),
i.e. as the system approaches the magnetic transition point.
This feature reveals that weakly magnetic adatoms or even non-magnetic
adatoms close to the transition point can host long-living 
longitudinal excitations~\cite{PhysRevLett.119.017203}.
In the following,
we first focus on quantitatively 
analyzing the ingredients of Eq. (\ref{eq:Tlong})
and subsequently turn to $T_{\parallel}$ itself.

\begin{figure}[t]
\includegraphics[width=\columnwidth]{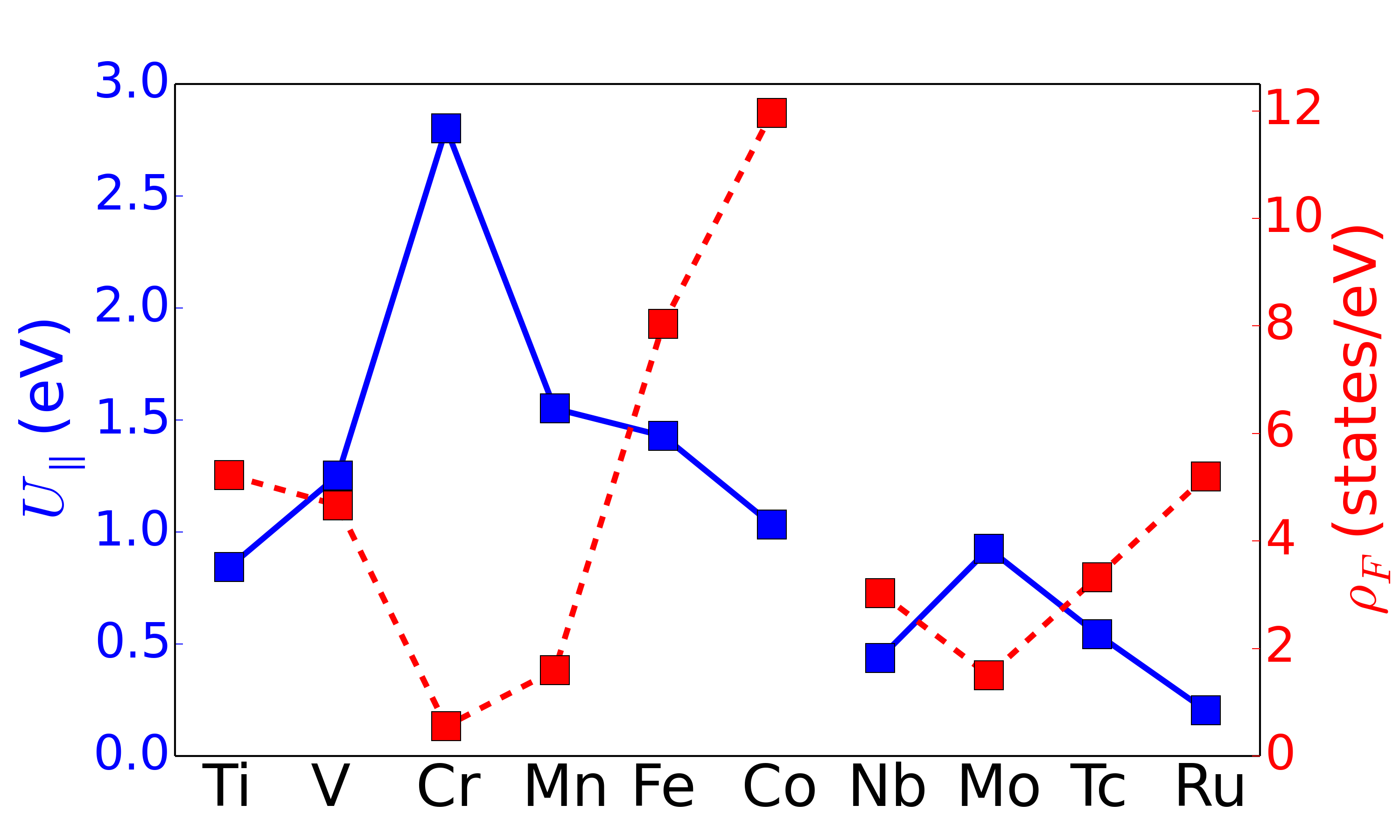}
\caption{(color online) 3\textit{d} and 4\textit{d} magnetic adatoms deposited on Ag(100). 
Solid (blue) and dashed (red) lines show the 
calculated values for $U_{\parallel}$ (from Eq. (\ref{eq:U_long})) and 
$\rho_{F}$
whose corresponding ordinate axes are placed on the left (blue) and right (red)
of the graph, respectively. Note that lines are broken in order to separate
3\textit{d} from 4\textit{d} elements.
}
\label{fig:Udotrho}
\end{figure}

In order to compute reliable values for the
kernel $U_{\parallel}$, we make use of 
the static limit of Eq. (\ref{eq:susc-long}), from which
\bek\label{eq:U_long}
U_{\parallel} = \rho_{F}^{-1}-\chi^{-1}_{\parallel}(0).
\ek
We note that $\chi_{\parallel}(0)$ can be calculated by a standard ground state DFT calculation
with a static magnetic field $\Delta B$ via $\chi_{\parallel}(0)=\Delta m /\Delta B$,
with $\Delta m$ the corresponding self-consistent change of the magnetic moment~\cite{kubler_theory_2009}. 
In Fig. \ref{fig:Udotrho} we show the calculated values of $\rho_{F}$ and $U_{\parallel}$
for  several 3\textit{d} and 4\textit{d} adatoms deposited on
Ag(100). The most important message exposed by this figure is the large variation
of $U_{\parallel}$ among different elements; while  $U_{\parallel}\lesssim 0.5$ eV
for most 4\textit{d} elements, $U_{\parallel}\gtrsim 1.5$ eV for various 3\textit{d} elements,
reaching a maximum of one order of magnitude difference between Ru and Cr. 
A second important feature revealed by Fig. \ref{fig:Udotrho} 
is the distribution of $U_{\parallel}$ within each \textit{d}-shell, whereby
it is smallest at the ends of the row  
--- case of Ti and Co among 3\textit{d}, Nb and Ru among 4\textit{d} --- 
and highest in the middle of the row 
--- case of Cr and Mn among 3\textit{d}, Mo and Tc among 4\textit{d} ---,
yielding an approximate inverted V-shape.
We note that $\rho_{F}$ in Fig. \ref{fig:Udotrho} shows the opposite behavior, 
i.e. it is minimum for Cr and maximum for Co and Ru.
This is consistent with Eq. (\ref{eq:U_long}), although we note strong deviations
from the $U_{\parallel}\propto \rho_{F}^{-1}$ relationship 
(see in particular the case of Cr), 
revealing the importance
of the term $\chi^{-1}_{\parallel}(0)$ in Eq. (\ref{eq:U_long}).

\begin{figure*}[t]
\subfloat[\label{fig:3d}]{%
  \includegraphics[width=.45\linewidth]{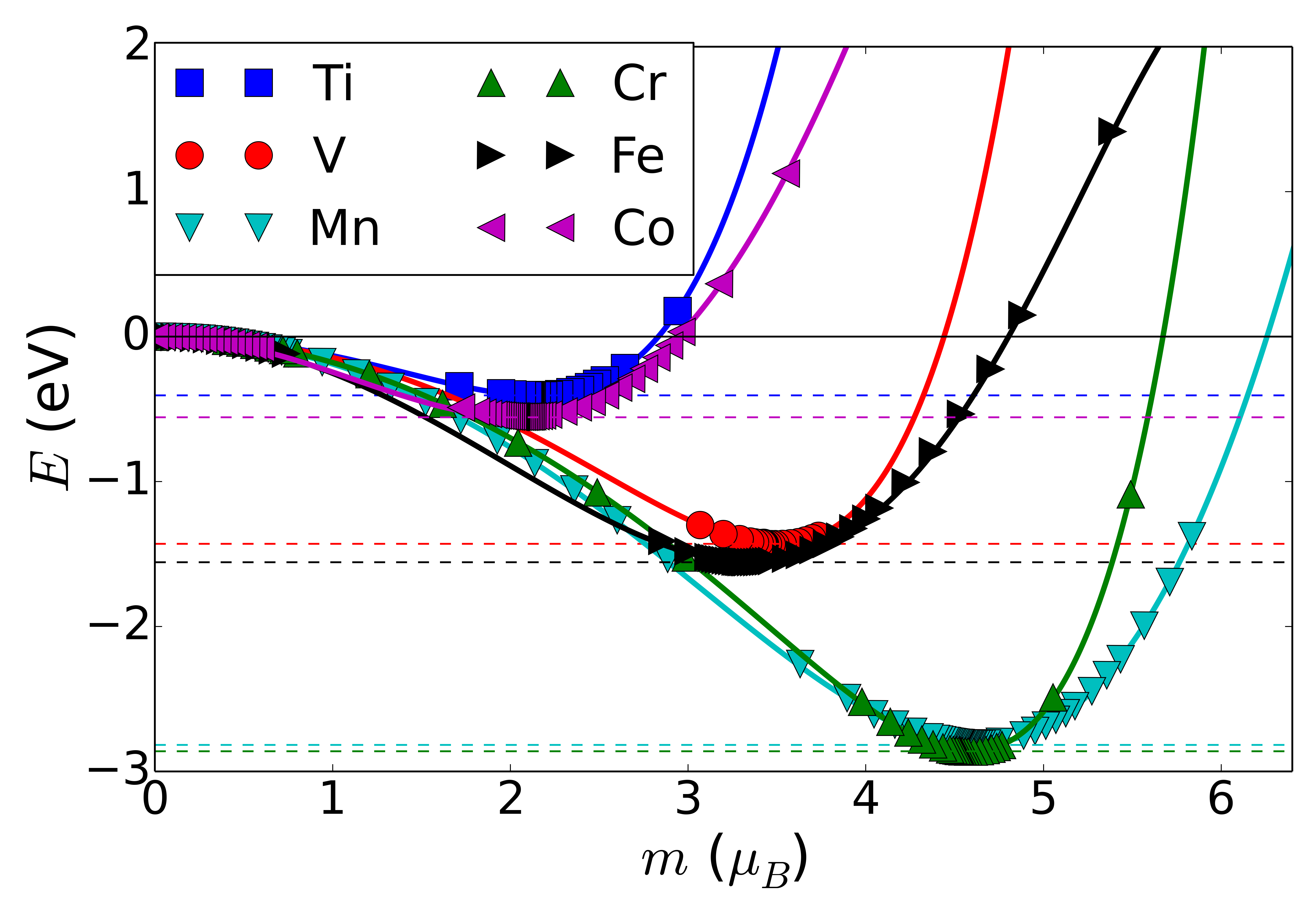}%
}\hfill
\subfloat[\label{fig:4d}]{%
  \includegraphics[width=.45\linewidth]{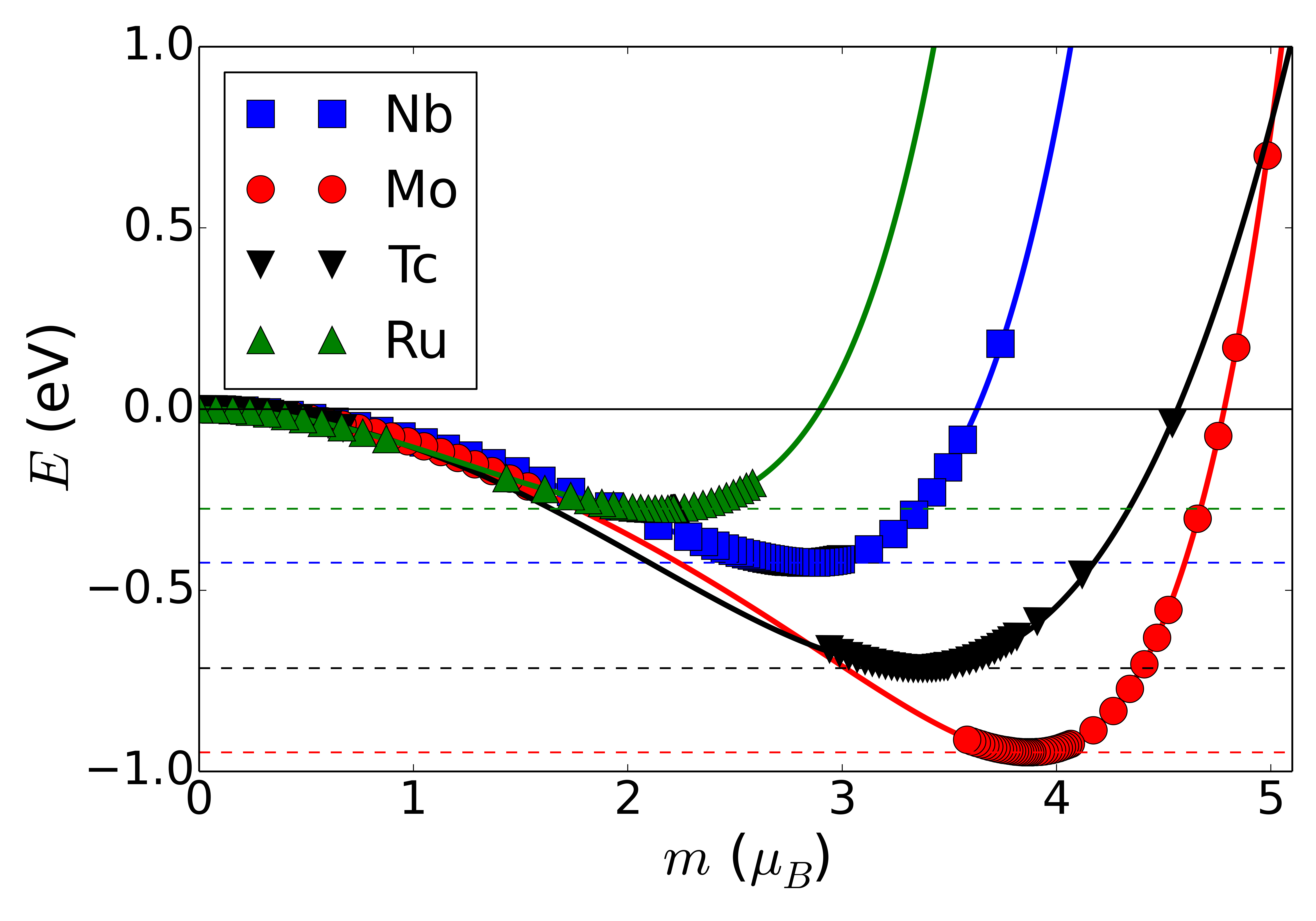}%
}
\caption{(color online) (a) and (b), calculated energy as a function of the adatom's magnetic moment for 
3\textit{d} and 4\textit{d} transition metal elements, respectively. DFT calculations are denoted by markers,
while the solid lines are a fit to the calculations using 
$E=\sum_{i=2}^{8} a_{i}m^{i}$ with only even powers.
The horizontal dashed lines 
mark the minimum energies for different adatoms.}
\label{fig:EM}
\end{figure*}

The inverse of the static spin-susceptibility  
$\chi^{-1}_{\parallel}(0)$ is closely connected to the magnetic equation of state,
i.e. the dependence of the energy $E$ as a function of the magnetic moment $m$,
via~\cite{kubler_theory_2009},
\bek\label{eq:E(m)}
\chi^{-1}_{\parallel}(0)= \dfrac{\partial^{2}E(m)}{\partial m^{2}}\Big|_{m=m_{0}}<0.
\ek
%with $m_{0}$ the ground state magnetic moment. 
In essence, the magnetic equation of state informs about how
stable the magnetic solution is in comparison to the non-magnetic one.
We have calculated $E(m)$ for the set of adatoms considered in Fig \ref{fig:Udotrho}
by employing DFT for fixed magnetic fields~\cite{kubler_theory_2009}. The results are shown in 
Fig. \ref{fig:EM}, which
%for several 3\textit{d} and 4\textit{d} adatoms deposited on Ag(100). 
reveals that the energy difference between the magnetic and non-magnetic state,
\bek
\Delta E = E(m=0) - E(m=m_{0}),
\ek
is of the order of eV and can largely vary for different adatoms.
Importantly, our calculations show that 3\textit{d} adatoms overall have a substantially 
larger $\Delta E$ than 4\textit{d} adatoms;
for Cr, for instance, $\Delta E\sim 3$ eV, while for Ru $\Delta E\sim 0.25$ eV.
Therefore, DFT predicts most 3\textit{d} adatoms to be magnetically more stable than 4\textit{d} ones,
as expected.
Furthermore, given that Eq. (\ref{eq:E(m)}) together with Eq. (\ref{eq:U_long}) 
relates the XC kernel to the second derivative of the
equation of state at $m_{0}$, one can establish an approximate connection between
the depth of the minimum of $E(m)$ and the value of $U_{\parallel}$,
as it is visible from the comparison of Figs. \ref{fig:EM} and \ref{fig:Udotrho};
the deeper the minimum, the larger $U_{\parallel}$.

We note that following the above procedure, one can also extract the 
kernel in the non-magnetic ground state,  
i.e. the so-called Stoner XC parameter $I_{xc}$~\cite{janak_uniform_1977}.
This can be achieved by considering the curvature of $E(m)$ not at $m=m_{0}$ but at
$m={0}$, as well as using the non-magnetic DOS in Eq. (\ref{eq:U_long})
instead of the magnetic one. As it is clearly visible from Fig. \ref{fig:EM}, 
the curvature is very different at $m={0}$ and $m=m_{0}$. 
Furthermore, $\rho_{F}$ can also strongly vary from a magnetic to a non-magnetic calculation.
As a consequence, the distribution of $I_{xc}$ along the transition metal series
first reported by Janak in Ref.~\onlinecite{janak_uniform_1977}
is very different
to that of $U_{\parallel}$ illustrated in Fig. \ref{fig:Udotrho}.

\begin{figure}[b]
\includegraphics[width=\columnwidth]{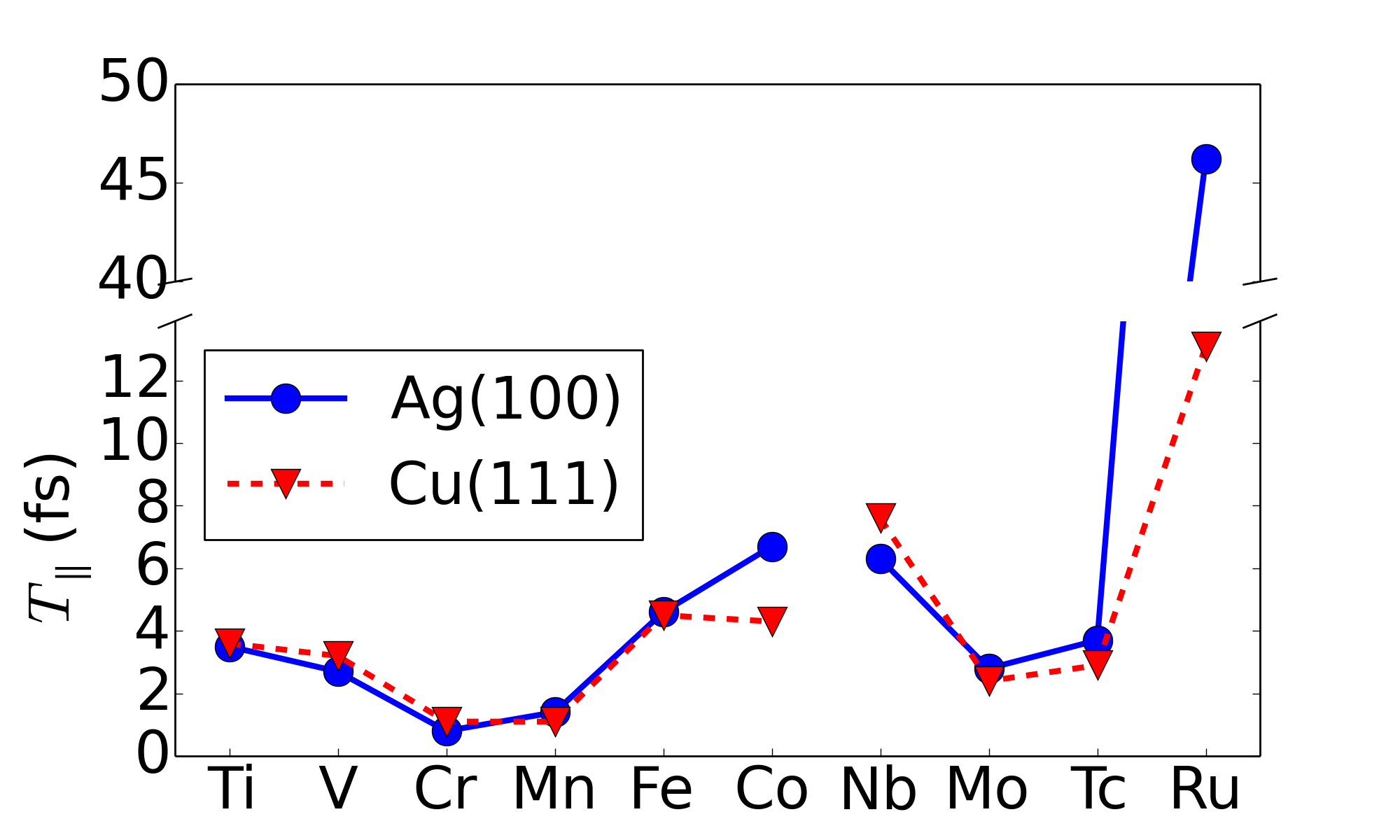}
\caption{(color online) Calculated values for the longitudinal relaxation time 
$T_{\parallel}$ for magnetic 3\textit{d} and 4\textit{d} 
    adatoms on Ag(100) (blue circles) and Cu(111) (red triangles).
}
\label{fig:Tparallel}
\end{figure}

Having analyzed the properties of
$\rho_{F}$ and $U_{\parallel}$, we next focus on the longitudinal relaxation time $T_{\parallel}$.
The values calculated from Eq. (\ref{eq:Tlong})
are plotted in Fig.~\ref{fig:Tparallel} for 3\textit{d} and 4\textit{d} adatoms deposited on Ag(100) and Cu(111). 
$T_{\parallel}$ is of the order of a few fs in all cases, being overal
slightly larger for 4\textit{d} than 3\textit{d} adatoms, while the choice of substrate
does not substantially affect it. 
Within each \textit{d}-shell, $T_{\parallel}$ is largest at the ends of the row 
while it is minimum for the half filled elements, thus resembling
the behavior of $\rho_{F}$ (compare Figs. \ref{fig:Udotrho} and \ref{fig:Tparallel}).
Ru on Ag(100) has the highest value of 
$T_{\parallel}\sim 50$ fs, mainly as a consequence of the denominator of 
Eq. (\ref{eq:Tlong}) being closer to zero than in other elements. 
In contrast,
Cr and Mn have $T_{\parallel}\sim 1$ fs in both substrates, i.e.
nearly two orders of magnitude less than the aforementioned example.
As a general feature, we note that the order of magnitude of $T_{\parallel}$ 
is settled by the energy scale of the problem:
all quantities involved in Eq. (\ref{eq:Tlong}) are of the order of eV,
whose corresponding time scale is fs.
Therefore, the longitudinal relaxation of the spin analyzed in this work is extremely fast.
The physical reason is the large exchange splitting dominating the relaxation process,
which makes it energetically very expensive to modify the length of the moment
due to the high energies involved.

\subsubsection{Connection to experimental measurements}
\label{subsubsec:long}

Let us next consider the experimental 
scenario regarding the measurement of the
longitudinal spin relaxation time.
For this, we first note that in a experiment,
several different mechanisms can contribute to this relaxation process,
whose overall relaxation time is generally denoted as $T_{1}$.
In this context, $T_{\parallel}$ calculated here is a particular
contribution to $T_{1}$, which may include further contributions depending on the
physical processes taking place. 
To the best of our knowledge, the first experimental 
technique that measures $T_{1}$  in magnetic single adatoms was developed by
Loth and co-workers~\cite{loth_measurement_2010}.
Within this STM-based technique, the spin relaxation time was measured 
by monitoring the decay of electrons in excited states after the application of an 
all-electronic pump-probe scheme~\cite{loth_measurement_2010}.
It is noteworthy that this scheme has so far only been applied 
to adatoms deposited on semi-insulating substrates, which  are  
close to the atomic limit.
The original work by Loth and co-workers measured $T_{1}\sim90$ ns for a Fe-Cu dimer on
$\text{Cu}_{2}$Ni/Cu(100)~\cite{loth_measurement_2010}. 
A subsequent work by Rau and co-workers measured  $T_{1}\sim200$ $\mu$s
for a single Co atom on MgO/Ag(100)~\cite{rau_reaching_2014}.
Lastly, Baumann and co-workers reported 
$T_{1}\sim90$ $\mu$s for a single Fe atom on MgO/Ag(100)~\cite{baumann_electron_2015},
while in a recent work of Paul and co-workers on the same system,~\cite{paul_control_2017} 
the value of $T_{1}$ was enhanced up to the ms regime by fine tuning
external conditions such as the height of the STM tip. 
To conclude, we note that the reported 
time resolution of the measuring technique employed in the above experiments 
ranges between few ns to hundreds of ps.

\begin{figure}[t]
\includegraphics[width=\columnwidth]{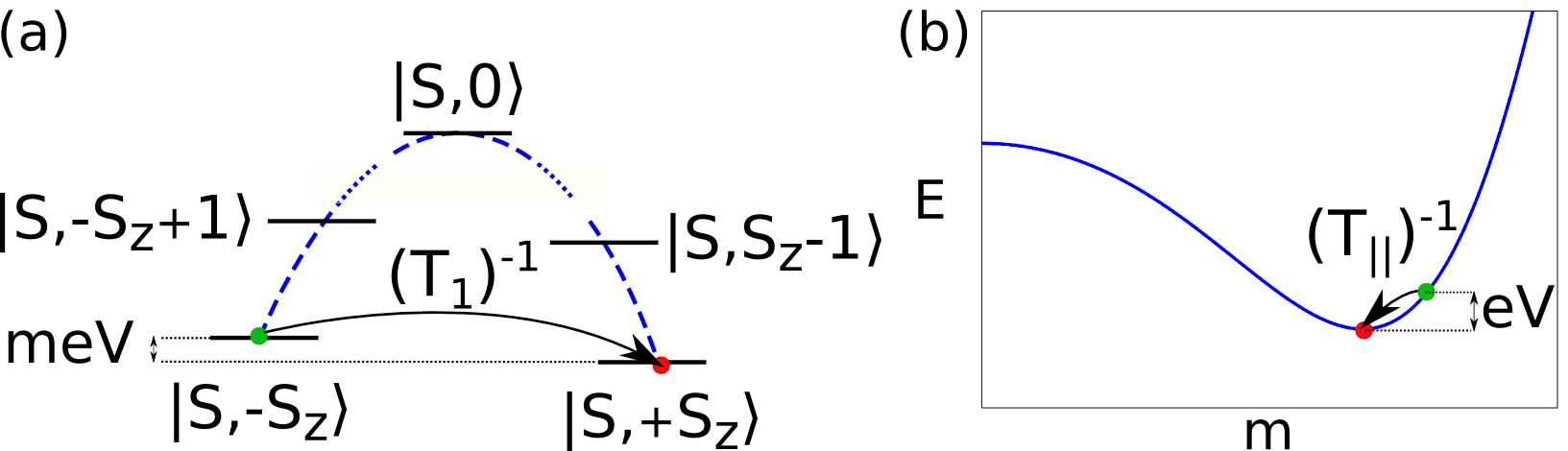}
\caption{(color online) Schematic illustration
of two different processes contributing to 
the spin relaxation time.
Green and red dots respectively represent the excited and ground states.
(a) illustrates the prototypical energy diagram  
used to describe the experiments of 
Refs.~\onlinecite{loth_measurement_2010,rau_reaching_2014,baumann_electron_2015,paul_control_2017}.
The initially degenerate two ground states of maximum $S_{z}$ become non-degenerate
by energy $\sim$ meV under externally applied magnetic fields of 1-10 T.
Transitions between these two states then determine $T_{1}$,
which take place via quantum tunneling. 
(b) schematically describes the 
excitations contributing to $T_{\parallel}$
considered in this work. These take place in a continuum energy landscape of 
order eV and are driven by direct spin-conserving electron-hole transitions.
}
\label{fig:E-diagram}
\end{figure}

All the above measured values of the spin relaxation time 
are several orders of magnitude larger 
than the values of order fs  that we have calculated
in this work for $T_{\parallel}$ (see Fig.~\ref{fig:Tparallel}).
Let us first note that all of the above experiments are performed under
externally applied static magnetic fields that range between 1 T and 10 T. This, in turn,
breaks the degeneracy of the spin ground states~\cite{loth_measurement_2010},  
a situation that is commonly modeled by 
a shifted discrete energy diagram as the one shown in Fig. \ref{fig:E-diagram}(a). 
We note that excitations within such a diagram 
are not allowed to change the length of the spin moment
(spin quantum number $S$ in this context), 
but only its projection (magnetic quantum number $S_{z}$).
Therefore, the main spin relaxation process contributing to $T_{1}$
within such an scheme involves transitions between the two non-degenerate states
with same $S$ but opposite $S_{z}$ (see Fig. \ref{fig:E-diagram}(a)).
We note that their energy separation is of order meV, hence much smaller than the 
excitations of order eV involved in the change of the spin magnetic moment size
considered for 
our calculation of $T_{\parallel}$, as schematically depicted in Fig.  \ref{fig:E-diagram}(b).
On top of that, given that direct transitions between the two non-degenerate 
states of Fig.  \ref{fig:E-diagram}(a) are virtually inexistent, 
spin relaxation in these conditions is  driven by quantum tunneling processes,
which are intrinsically much slower than the direct transitions considered in this work. 
These two considerations explain why the spin relaxation time  measured
under the mentioned experimental conditions
is several orders of magnitude larger than the values of $T_{\parallel}$ obtained
in this work.

It is apparent that, in order to experimentally 
access the dynamics encoded into  $T_{\parallel}$, 
a measuring scheme based on ultrafast techniques that modify the
length of the spin magnetic moment is required. 
Considering the technological developments within 
STM measuring techniques~\cite{kruger_attosecond_2011,cocker_ultrafast_2013,cocker_tracking_2016}, 
accessing the fs time scale of magnetic adatoms seems to be a reasonable goal
for the near future by, e.g. using ultrafast laser pulses, 
a breakthrough that would allow to monitor the ultrafast spin-dynamics
analyzed in this work.

\subsection{Transverse component}
\label{subsec:trans}

Unlike the longitudinal component, the transverse spin-susceptibility 
and associated spin-excitations of single adatoms have been thoroughly studied 
from first principles in, \textit{e.g.}, Refs.
~\onlinecite{lounis_dynamical_2010,lounis_theory_2011,dias_relativistic_2015,PhysRevB.91.104420,PhysRevB.89.235439,ibanez-azpiroz_zero-point_2016}. 
The general form for the adatom's 
enhanced transverse spin-susceptibility (see Eq. (\ref{eq:dm+_w})) 
in the 
TDDFT scheme is
\bek\label{eq:chi-tr}
\chi_{\pm}(\omega)=\dfrac{\chi_{\pm}^{\text{KS}}(\omega)}{1-U_{\perp}\chi^{\text{KS}}_{\pm}(\omega)}.
\ek
Above, $U_{\perp}$ is the transverse XC kernel
treated in the adiabatic local spin-density approximation~\cite{katsnelson_magnetic_2004},
while $\chi_{\pm}^{\text{KS}}$ denotes the transverse KS
spin-susceptibility.
Despite the formal similarity between Eq. (\ref{eq:susc-long}) for $\chi_{\parallel}(\omega)$
and Eq. (\ref{eq:chi-tr}) for $\chi_{\pm}(\omega)$, the underlying physics behind both 
expressions is very different. While the former contains 
excitations that modify the spin density, the latter describes 
damped precessional motion of the spin moment~\cite{white_quantum_2007}.
This motion, in turn, is described by 
the imaginary part of the enhanced spin-susceptibility of Eq. (\ref{eq:chi-tr}),
$\Im\chi_{\pm}(\omega)$, 
which gives access to the density of transverse spin-excitations of  
single adatoms ~\cite{lounis_dynamical_2010,lounis_theory_2011,dias_relativistic_2015,PhysRevB.91.104420,PhysRevB.89.235439,ibanez-azpiroz_zero-point_2016}.

\begin{figure}[t]
\includegraphics[width=0.4\textwidth]{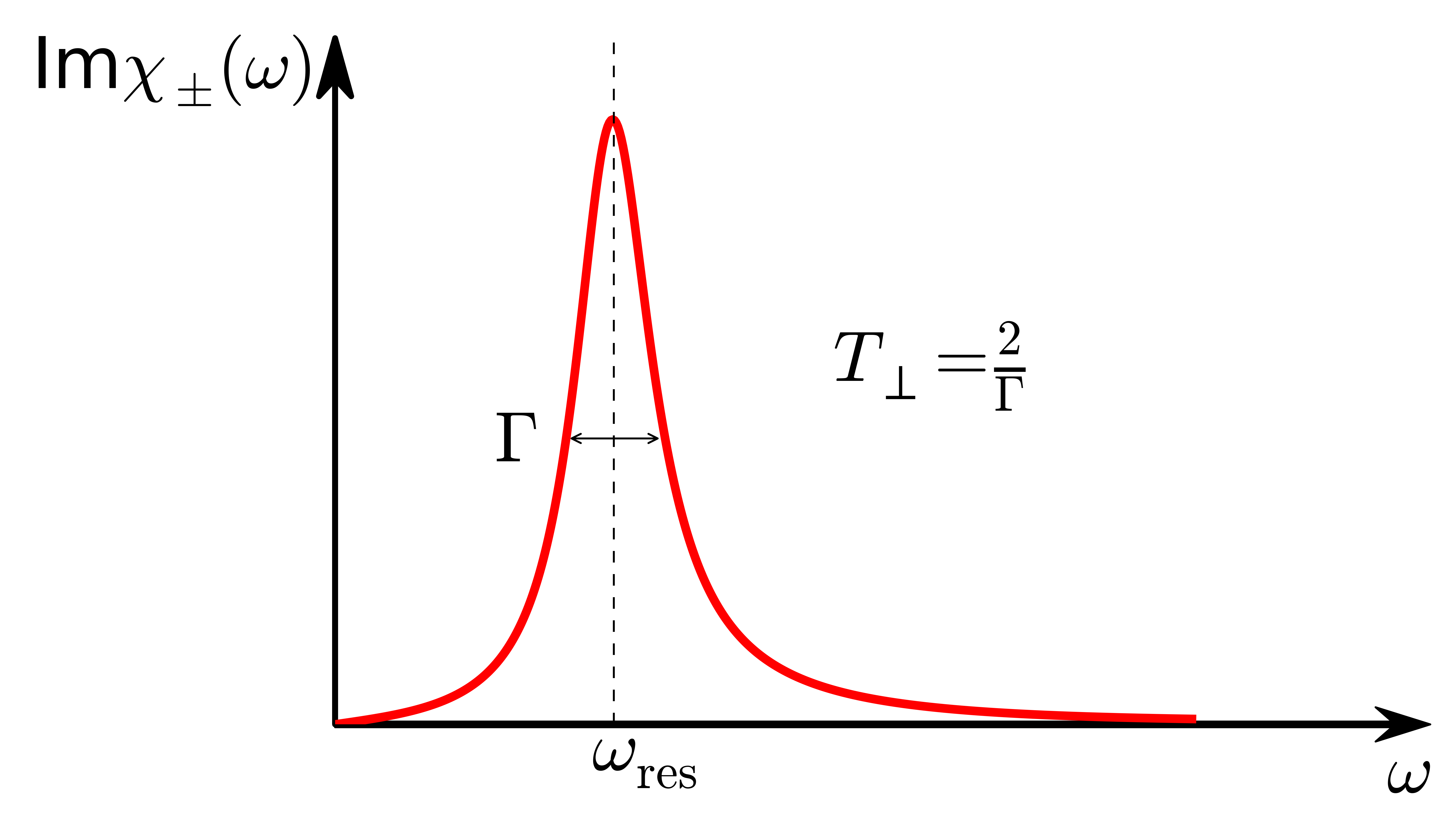}
\caption{(color online) Schematic illustration of a transverse spin-excitation of a single adatom contained
in $\Im\chi_{\pm}(\omega)$ (see Eq. (\ref{eq:chi-tr})). The resonance frequency, 
$\omega_{\text{res}}$, and the width, $\Gamma$, are indicated in the figure.
}
\label{fig:t-trans}
\end{figure}

The characteristic form of the spin-excitation hosted by $\Im\chi_{\pm}(\omega)$ is 
illustrated in Fig. \ref{fig:t-trans} and
is characterized  by two main quantities. The first one is
its resonance frequency, $\omega_{\text{res}}$, 
a fundamental property related to the magnetic anisotropy energy that is
ultimately determined by SOC~\cite{dias_relativistic_2015}.
The second main quantity is the width of the spin-excitation, $\Gamma$, 
which is  proportional to the hybridization of the adatom's electrons with 
the substrate~\cite{lounis_dynamical_2010,lounis_theory_2011}
(see also Ref.~\onlinecite{ternes_spin_2015} for 
model Hamiltonian point of view).
As shown in Ref.~\onlinecite{PhysRevB.91.104420},
the main contribution of the hybridization to $\Gamma$ is proportional
to the electron-hole excitations of opposite spin channel, 
$n_{e\text{-}h}^{\prime} = \pi(\rho_{F,\uparrow}\cdot\rho_{F,\downarrow})$:
\bek\label{eq:hybr-trans}
\Gamma\simeq \dfrac{n_{e\text{-}h}^{\prime}}{\text{Re}Q}  \omega_{\text{res}},
\ek
with $Q=\partial\chi^{\text{KS}}_{\pm}(\omega)/\partial\omega\Big|_{\omega=0}$. 
We note that the order of magnitude of $\omega_{\text{res}}$ ranges between 
$10^{-2}-1$ meV while
$n_{e\text{-}h}^{\prime}/\text{Re}Q$ is a unitless fraction that is typically of order 
unity.

\begin{figure}[t]
\includegraphics[width=\columnwidth]{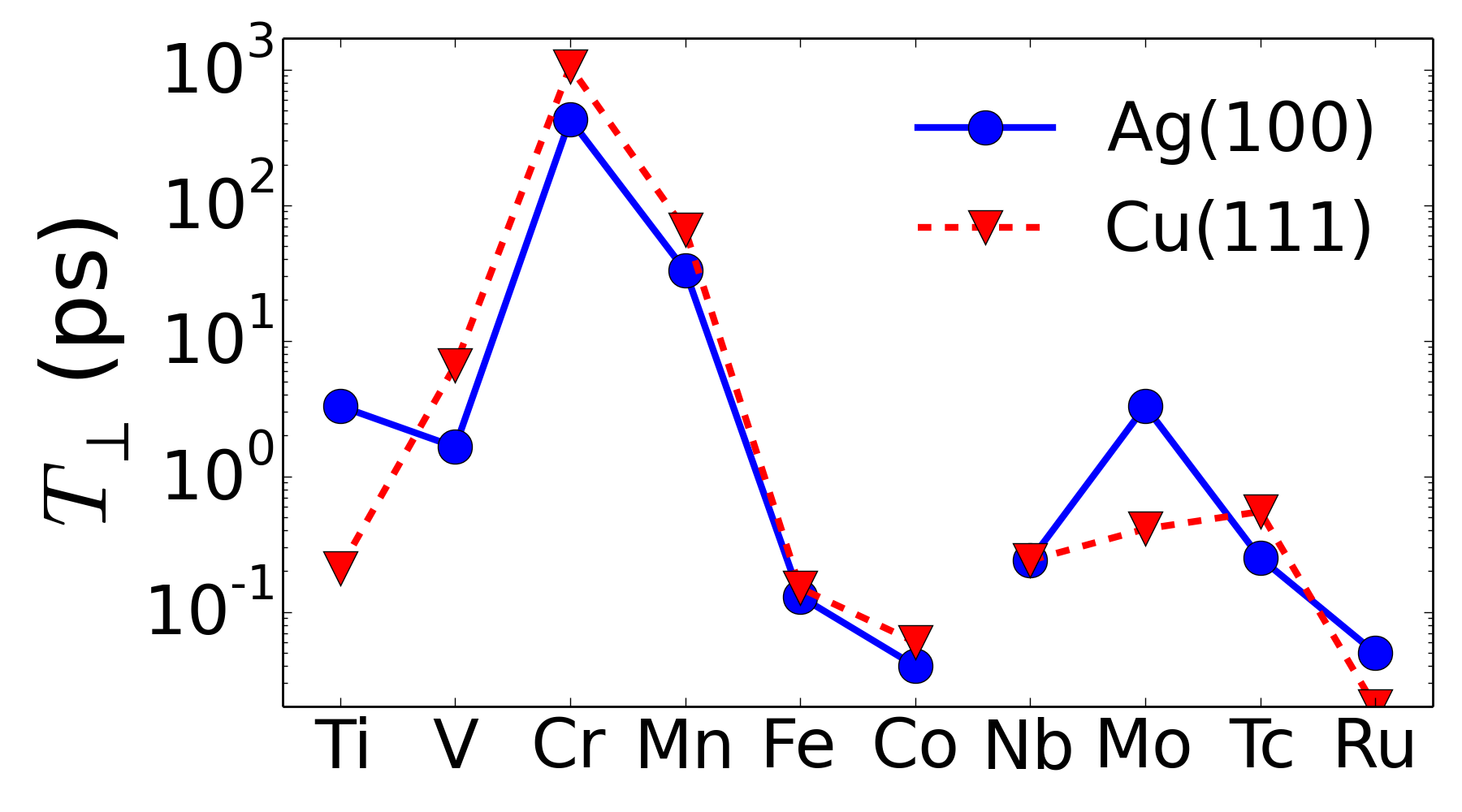}
\caption{(color online) Calculated values for the transverse 
relaxation time $T_{\perp}$ for magnetic 3\textit{d} and 4\textit{d} 
    adatoms on Ag(100) (blue circles) and Cu(111) (red triangles). 
    Note the logarithmic scale in the y axis.
}
\label{fig:Tperp}
\end{figure}

Importantly, a finite width corresponds to a finite 
damping of the precessing magnetic moment 
and is thus directly linked to
the transverse spin relaxation time (see Appendix \ref{appendix:LLG}
and Eq. (\ref{eq:hybr-trans})): 
\bek\label{eq:Tperp-Gamma}
T_{\perp}=\dfrac{2}{\Gamma}\propto \left(n_{e\text{-}h}^{\prime}\right)^{-1}.
\ek
We note that, while $T_{\parallel}$ in Eq. (\ref{eq:Tlong})
is directly proportional to the density of 
spin-conserving electron-hole excitations $n_{e\text{-}h}$,
$T_{\perp}$ above is inversely proportional to the 
spin-flip counterpart $n_{e\text{-}h}^{\prime}$.

Using the TDDFT formalism 
developed in Refs.~\onlinecite{lounis_dynamical_2010,lounis_theory_2011,dias_relativistic_2015}, we 
have calculated $T_{\perp}$ from the spin-excitation width
for various 3\textit{d} and 4\textit{d} magnetic adatoms deposited on the metallic substrates
Ag(100) and Cu(111); calculated values are shown in Fig.~\ref{fig:Tperp}. 
One notes that the variation of $T_{\perp}$ among adatoms is somewhat larger than that of 
$T_{\parallel}$, shown in Fig.~\ref{fig:Tparallel}. 
This is a consequence of the large variation of the width $\Gamma$ 
of atomic spin-excitations, which can range from 10$^{-2}$ meV to 
few meV~\cite{dias_relativistic_2015,ibanez-azpiroz_zero-point_2016},
i.e. nearly three orders of magnitude change.
This, in turn, can be linked to the electronic  DOS at the Fermi level via Eq. \ref{eq:hybr-trans};
adatoms where the DOS peak of the \textit{d} states lies close to the Fermi level, case of  
Ti, V, Fe and Co in Fig. \ref{fig:DOS}, tend to be much more hybridized than those 
where only the tail of the DOS peak lies at the Fermi level, case of Cr and Mn in Fig. \ref{fig:DOS}.
In this way, Cr and Mn acquire large relaxation times of  $T_{\perp}\sim 10^{1}$--$10^{2}$ ps,
while strongly hybridized adatoms such as Co, 
Nb and Ru
have $T_{\perp}\sim 10^{-1}$--$10^{-2}$ ps.

To conclude, let us note that the trend of $T_{\perp}$
within each \textit{d}-shell row is opposite to that shown by $T_{\parallel}$ (see Fig.~\ref{fig:Tparallel}).
This comes as a consequence of the dependence on the density of electron-hole excitations,
with $T_{\parallel}\propto n_{e\text{-}h}$ (see Eq. (\ref{eq:Tlong}))
and $T_{\perp}\propto (n_{e\text{-}h}^{\prime})^{-1}$ (see Eq. (\ref{eq:hybr-trans})).
In fact, $T_{\parallel}$ and $T_{\perp}$
can be formally related to each other considering the relation between
the spin-conserving and spin-flip electron-hole contributions, namely
$\rho_{F}^{2}/2=n_{e\text{-}h}+n_{e\text{-}h}^{\prime}$. 
From Eq. (\ref{eq:Tlong}) for $T_{\parallel}$ 
and Eq. (\ref{eq:hybr-trans}) for $T_{\perp}$ one can then infer
the following expression:
\bek\label{eq:relation-Tpara-Tperp}
\dfrac{\rho_{F}^{2}}{2}\simeq
T_{\parallel}\dfrac{U_{\parallel}\rho_{F}-1}{U_{\parallel}}
+2T_{\perp}^{-1}\dfrac{\text{Re}Q}{\omega_{\text{res}}}.
\ek
The unifying concept behind the above relationship between $T_{\parallel}$ and $T_{\perp}$
is the hybridization of substrate electrons with the \textit{d}-states of the transition metal 
adatoms, which in essence gives rise to a finite total $\rho_{F}$. 
However, despite the formal relationship,
the fact that 
$T_{\parallel}$ and $T_{\perp}$ in Eq. \ref{eq:relation-Tpara-Tperp}
have fundamentally different prefactors
makes the time scale of the two relaxation constants
differ by nearly three orders of magnitude.

\subsubsection{Connection to experimental measurements and 
a comment on nomenclature}
\label{subsubsec:long}

Next, we consider several experimental measurements of 
spin-excitation lifetimes of different single adatoms and 
connect them to our work.
The lifetime of an atomic spin-excitation can be experimentally accessed  
from the width of the step observed in IETS $\text{d}I/\text{d}V$ measurements, which
provides a measure of $\Gamma$. 
Given that the energy resolution of this technique is 10$^{-1}$ meV at best~\cite{PhysRevLett.111.157204},
the longest lifetimes that can be inferred following this procedure
are of order 10 ps (see Eq. \ref{eq:Tperp-Gamma}).  
These type of experiments can measure adatoms deposited on both 
metallic and semi-insulating substrates;
as a general trend, the latter induce a larger lifetime than the former 
due to a far smaller electronic hybridization.
We  begin by considering Ref. \onlinecite{PhysRevLett.106.037205}, where 
Khajetoorians and co-workers estimate the spin-excitation lifetime  
of a Fe adatom deposited on metallic Cu(111) to be 
0.2 ps, in very good quantitative 
agreement with our calculated value $T_{\perp}=0.15$ ps for the same system
(see  Fig.~\ref{fig:Tperp}).
Noteworthily, when the same atom is deposited on metallic Pt(111), 
the measured lifetime  is increased by
nearly an order of magnitude~\cite{PhysRevLett.111.157204}.
We note that we have found a similar variation 
between the two substrates  considered in this work
for the elements Ti, V, Cr and Mo, as it can be checked in Fig.~\ref{fig:Tperp}. 
Focusing next on the semi-insulating $\text{Cu}_{2}$Ni/Cu(100) substrate,
a lower bound of  $\sim$10 ps has been experimentally estimated for
Fe~\cite{hirjibehedin_large_2007,loth_njp}, Mn~\cite{loth_njp} and Co~\cite{otte_role_2008}
adatoms, although it is possible that the actual lifetimes are substantially larger.
In fact, our calculations on Cr and Mn, which are the elements with smallest hybridization
and thus the ones closest to the semi-insulating limit, show that  $T_{\perp}$ can reach up to
10$^{3}$ ps (see Fig.~\ref{fig:Tperp});
hence, it is not unlikely that the lifetimes of the aforementioned  
adatoms on $\text{Cu}_{2}$Ni/Cu(100) could be of the same order of magnitude.
Last, it is worth noting the case of Co on MgO~\cite{rau_reaching_2014},
which, despite being a semi-insulating substrate, yields a relatively short 
spin-excitation lifetime of $\sim$ 0.5 ps, i.e. a common value for
adatoms deposited in metallic substrates analyzed in this work (see Fig. \ref{fig:t-trans}).

To conclude this section, 
we note that the convention followed in the standard literature to 
denote the transversal relaxation time is $T_{2}$
(see, \textit{e.g.}, Refs. \onlinecite{white_quantum_2007,canet_relaxation_2006}).
We have noticed, however, that the relaxation time associated to atomic spin-excitations
has in some cases been named as a $T_{1}$-like term;
see, \textit{e.g.}, 
the review by Delgado and Fern{\'a}ndez-Rossier~\cite{delgado_spin_2017}.
The authors of this review use $T_{2}$ to denote another relaxation mechanism named as
adiabatic decoherence. 
We have included a brief discussion on this nomenclature
issue in Appendix \ref{appendix:BR}.

\section{Conclusions}
\label{sec:discussion}

In conclusion, we have presented a systematic \textit{ab initio}
investigation of longitudinal and transverse spin relaxation times 
of  magnetic single adatoms deposited on metallic substrates. 
Our analysis has yielded as a main
result the fact that the longitudinal spin relaxation process  
of  single adatoms is much faster
than the transverse one, i.e., $T_{\perp}\gg T_{\parallel}$.
This, in turn, comes as a consequence of the energy scale of the corresponding
processes; eV for $T_{\parallel}$, meV for $T_{\perp}$. Importantly,  the two processes 
are triggered by different mechanisms:
while $T_{\parallel}$ is driven by spin-conserving 
excitations that change the spin-density, $T_{\perp}$ 
depends on the atomic spin-flip spin-excitations that induce   
the precessional motion.
The comparison of our results with available 
experimental measurements  shows that 
the relaxation times extracted from 
inelastic scanning tunneling spectroscopy $\mathrm{d}I/\mathrm{d}V$ 
curves show overal
the same order of magnitude as $T_{\perp}$ 
and agree remarkably well in
specific cases such as Fe on Cu(111)~\cite{PhysRevLett.106.037205}.
Regarding the measurement of $T_{\parallel}$, we have argued  that,
although currently available techniques
cannot monitor the femtosecond regime of magnetic single adatoms, 
it is reasonable that this can be achieved in the near future 
\textit{e.g.} by
employing STM-integrated ultrafast laser schemes~\cite{kruger_attosecond_2011,cocker_ultrafast_2013,cocker_tracking_2016},
thus giving access to the ultrafast spin-dynamics
described in this work.

\section{Acknowledgments}
The authors are very grateful to F. Guimar\~{a}es
for fruitful discussions on the LLG model.
This work has been supported by the Impuls und
Vernetzungsfonds der Helmholtz-Gemeinschaft Postdoc Programme 
and funding from the European Research Council (ERC) under 
the European Union's Horizon 2020 research and innovation 
programme (ERC-consolidator grant 681405 — DYNASORE).
The authors gratefully acknowledge the computing 
time granted by the JARA-HPC Vergabegremium and provided 
on the JARA-HPC Partition part of the supercomputer JURECA at Forschungszentrum J\"ulich.

\appendix

%%%%%%%%%%%%%%%%%%%%%%%%%%%%%%%%%%%%%%%%%%%%%%%%%%%%%%%%%%%%%%%%%%%%%%
%%%%%%%%%%%%%%%%%%%%%%%%%%%%%%%%%%%%%%%%%%%%%%%%%%%%%%%%%%%%%%%%%%%%%%
\section{Longitudinal Bloch equation}
\label{appendix:Bloch-long}

The Bloch equation for the longitudinal change 
of the magnetization under the effect
of a time-dependent perturbation $H_{1}(t)$ along the longitudinal direction 
can be written as~\cite{white_quantum_2007}
\begin{equation}
\label{eq:bloch-T1-t}
\dfrac{dm_{z}(t)}{dt}=\dfrac{\chi^{\text{Bl}}_{0}H_{1}(t)-m_{z}(t)}{T_{\parallel}},
\end{equation}
with $\chi^{\text{Bl}}_{0}$ a static spin-susceptibility.
The above equation describes how $m_{z}(t)$ comes back to equilibrium 
with a characteristic relaxation time $T_{\parallel}$
after being perturbed by $H_{1}(t)$. 
Using $f(t)=\int d\omega f(\omega)e^{-i\omega t}$
for both $m_{z}(t)$ and $H_{1}(t)$ we can write Eq. \ref{eq:bloch-T1-t} in frequency domain,
\begin{equation}
\begin{split}
\label{eq:bloch-T1-w}
&m_{z}(\omega)(-i\omega T_{\parallel} + 1)=\chi_{0}H_{1}(\omega) \Rightarrow \\
&
\dfrac{m_{z}(\omega)}{H_{1}(\omega)}\equiv \chi^{\text{Bl}}(\omega) =
\dfrac{\chi^{\text{Bl}}_{0}}{1-i\omega T_{\parallel}},
\end{split}
\end{equation}
where  $\chi^{\text{Bl}}(\omega)$ is the enhanced spin-susceptibility. 
The real and imaginary parts of the above equation read
\begin{eqnarray}
\label{eqs:bloch-re}
& \Re \chi^{\text{Bl}}(\omega) =  \dfrac{\chi^{\text{Bl}}_{0}}{1+(\omega T_{\parallel})^{2}}, \\
\label{eqs:bloch-im}
& \Im \chi^{\text{Bl}}(\omega) =  \dfrac{\chi^{\text{Bl}}_{0}\omega T_{\parallel}}{1+(\omega T_{\parallel})^{2}}. 
\end{eqnarray}

Next, we consider the Taylor expansion of the KS spin-susceptibility
(see Eq. (\ref{eq:chi-KS-long}) in the main text),
\bek\label{eq:chi-KS-long-appendix}
\chi^{KS}_{\parallel}(\omega)\simeq \rho_{F}   -in_{e\text{-}h}\omega.
\ek
The first-order expansion coefficient 
$n_{e\text{-}h}=\pi(\rho^{2}_{F,\uparrow}+\rho^{2}_{F,\downarrow})/2$ has been calculated in the Supplemental
Material of Ref. \onlinecite{PhysRevLett.119.017203} .
Inserting $\chi^{KS}_{\parallel}(\omega)$ of 
Eq. (\ref{eq:chi-KS-long-appendix}) into the definition of the
TDDFT enhanced spin-susceptibility $\chi(\omega)$ 
(see Eq. (\ref{eq:susc-long}) of the main text), the imaginary part
$\text{Im}\chi_{\parallel}(\omega)$ can be cast in the following way,
\begin{equation}
\begin{split}
\label{eq:susc-T1-conn}
& \Im\chi_{\parallel}(\omega)=\dfrac{n_{e\text{-}h}\omega}
{(1-U_{\parallel}\rho_{F})^{2}}
\cdot\dfrac{1}
{1+\left(\dfrac{U_{\parallel}n_{e\text{-}h}\omega}{1-U_{\parallel}\rho_{F}}\right)^{2}} =\\ 
& \dfrac{\omega T_{\parallel}}
{(1-U_{\parallel}\rho_{F})U_{\parallel}}
\cdot\dfrac{1}
{1+(\omega T_{\parallel})^{2}}=
\dfrac{\chi(0)}{U_{\parallel}\rho_{F}}
\cdot\dfrac{\omega T_{\parallel}}
{1+(\omega T_{\parallel})^{2}},
\end{split}
\end{equation}
where in the last step we used the expression for the static spin-susceptibility
$\chi_{\parallel}(0) =\rho_{F}/(1-U_{\parallel}\rho_{F})$
and we defined 
\begin{equation}
T_{\parallel}=\dfrac{U_{\parallel}n_{e\text{-}h}}{1-U_{\parallel}\rho_{F}},
\end{equation}
which is the result quoted in the main text in Eq. (\ref{eq:Tlong}).

%%%%%%%%%%%%%%%%%%%%%%%%%%%%%%%%%%%%%%%%%%%%%%%%%%%%%%%%%%%%%%%%%%%%%%
%%%%%%%%%%%%%%%%%%%%%%%%%%%%%%%%%%%%%%%%%%%%%%%%%%%%%%%%%%%%%%%%%%%%%%
\section{Transverse relaxation within the Landau-Lifshitz-Gilbert equation}
\label{appendix:LLG}

We consider the Landau-Lifshitz-Gilbert (LLG) equation describing 
the damped precessional motion of
a magnetic moment placed in a static external magnetic field that has been 
perturbed by a time-dependent transverse magnetic field:
\begin{equation}\label{eq:LLG-eom}
\frac{\ud\VEC{m}}{\ud t} = -\gamma\,\VEC{m} \times \VEC{B}^{\text{ext}} + \eta\,\frac{\VEC{m}}{m_0} \times \frac{\ud\VEC{m}}{\ud t},
\end{equation}
with $\VEC{B}^{\text{ext}} = B_0\,\hat{\VEC{z}} + \textbf{b}(t)$
and $\textbf{b}(t)=\theta(t)\Delta B(\hat{\VEC{x}}+\hat{\VEC{y}})$, 
i.e. for $t < 0$ the static field points in the $z$-direction
while for $t > 0$ a small transverse component is switched on.
Note that the precession rate in Eq. (\ref{eq:LLG-eom}) 
is set by $\gamma$, while the relaxation is controlled by $\eta$, namely the damping term.
Let us assume $\VEC{m} = m_0\,\hat{\VEC{z}}$ for $t<0$.
Then, linearizing the LLG equation yields the following equation of motion for
the transverse components of the magnetization:
\begin{align}
  \frac{\ud m_x}{\ud t} &= -\gamma\,B_0\,m_y + \gamma\,m_0\,\Delta B- \eta\,\frac{\ud m_y}{\ud t} ,
  \label{eq:LLG-lin-x} \\
  \frac{\ud m_y}{\ud t} &= -\gamma\,m_0\,\Delta B + \gamma\,B_0\,m_x + \eta\,\frac{\ud m_x}{\ud t}  
  \label{eq:LLG-lin-y} .
\end{align}
Since the expected solution is a damped precession that relaxes towards the direction of the static magnetic field, we use the following ansatz corresponding to a circular precession %with shifted center
that decays in time with a transverse relaxation time $T_{\perp}$:
\begin{align}
  m_x(t) &= m_x(\infty) - A\,e^{-t/T_{\perp}}\cos(\omega_0 t)  , \\
  m_y(t) &= m_y(\infty) - A\,e^{-t/T_{\perp}}\sin(\omega_0 t) .
\end{align}
Plugging the above ansatz back into the LLG equation (\ref{eq:LLG-eom}) we get
\begin{equation}
\begin{split}
  &\frac{1}{T_{\perp}}\cos(\omega_0 t) + \omega_0\sin(\omega_0 t)= 
\frac{\gamma\,\big(-B_0\,m_y(\infty) + m_0\,\Delta B\big)}{A} \\  
 & \gamma\,B_0\sin(\omega_0 t)  
    -\eta\,\left(\frac{1}{T_{\perp}}\sin(\omega_0 t) - \omega_0\cos(\omega_0 t)\right) , \\
\end{split}
\end{equation}
\begin{equation}
\begin{split}
  &\frac{1}{T_{\perp}}\sin(\omega_0 t) - \omega_0\cos(\omega_0 t) = \frac{\gamma\,\big(B_0\,m_x(\infty) - m_0\,\Delta B\big)}{A}  \\
  & -\gamma\,B_0\,\cos(\omega_0 t) + \eta\left(\frac{1}{T_{\perp}}\cos(\omega_0 t) + \omega_0\sin(\omega_0 t)\right) .
\end{split}
\end{equation}
The above equations can only be satisfied if the coefficients 
in front of the time-dependent sines and cosines match.
We then have (both equations give the same pair of relations)
\bek
  &\dfrac{1}{T_{\perp}} = \eta\,\omega_0 ,\label{eq:LLG-Ttrans}\\
 & \omega_0 = \dfrac{\gamma\,B_0}{1+\eta^2} \label{eq:LLG-wmax},\\
& m_x(\infty) = m_y(\infty) = \dfrac{m_0}{B_0}\,\Delta B = \chi_{\perp}^{\text{LLG}}\,\Delta B,
\ek
where $\chi_{\perp}^{\text{LLG}}$ is the static spin-susceptibility and 
continuity of $m_x(t)$ and  $m_y(t)$ at $t = 0$ fixes $A = m_x(\infty)$.
Importantly, 
Eq. (\ref{eq:LLG-Ttrans}) shows that the transverse relaxation time is given by the product between
the damping term $\eta$ and the characteristic frequency $\omega_0$.

We next turn to calculate the transverse dynamic spin-susceptibility within the LLG model.
For this, we consider the following Fourier transforms,
\begin{equation}
  \VEC{b}(t) = \int\frac{\ud\omega}{2\pi}\,e^{-\iu\omega t}\,\VEC{b}(\omega) ,\;
  \VEC{m}(t) = \int\frac{\ud\omega}{2\pi}\,e^{-\iu\omega t}\,\VEC{m}(\omega).
\end{equation}
Inserting the above expressions into the linearized equations 
(\ref{eq:LLG-lin-x}) and (\ref{eq:LLG-lin-y}) we obtain in frequency space
\begin{align}
  -i\omega m_x(\omega) &= + \gamma m_0 b_y(\omega) + (i\eta \omega-\gamma B_0) m_y(\omega) , \\
  -i\omega m_y(\omega) &= -\gamma m_0 b_x(\omega) + (\gamma B_0- i\eta \omega) m_x(\omega).
\end{align}
The above can be simplified by considering the circular components 
$m_{\pm}=m_{x}\pm im_{y}$, yielding
\begin{equation}\label{eq:LLG-pre-chi}
\begin{split}
-i\omega m_{\pm}(\omega)&=\gamma_{\pm}\big(B_{0}m_{\pm}(\omega)-m_{0}b_{\pm}(\omega)\big), \Rightarrow \\
&\Lambda_{\pm}(\omega) m_{\pm}(\omega) = b_{\pm}(\omega),
\end{split}
\end{equation}
with $b_{\pm}(\omega)=b_{x}\pm ib_{y}$, $\gamma_{\pm}=\pm i\gamma/(1\mp i\eta)$ and
\bek\label{eq:LLG-lambda}
\Lambda_{\pm}(\omega) = \dfrac{1}{\gamma_{\pm}m_{0}}(i\omega+\gamma_{\pm}B_{0}).
\ek
It is apparent from Eq. (\ref{eq:LLG-pre-chi}) that 
the transverse spin-susceptibility can be obtained 
from the inverse of $\Lambda_{\pm}(\omega)$
defined above. After some algebra and picking the minus sign in Eq. (\ref{eq:LLG-lambda}) one obtains
\begin{equation}
\begin{split}
\chi^{\text{LLG}}_{\pm}(\omega)&=\big(\Lambda_{-}(\omega)\big)^{-1}  \\
&
=\dfrac{m_{0}\omega_{0}}{B_{0}}
\dfrac{-\omega+(1+\eta^{2})\omega_{0}+i\eta\omega}{(\omega-\omega_{0})^{2}+(\eta\omega_{0})^{2}}.
\end{split}
\end{equation}
The density of spin-excitations in the LLG model are thus described by a skewed Lorentzian in $\omega$:
\bek\label{eq:LLG-imchi}
\Im \chi^{\text{LLG}}_{\pm}(\omega)=
\dfrac{m_{0}\omega_{0}}{B_{0}}
\dfrac{\eta\omega}{(\omega-\omega_{0})^{2}+(\eta\omega_{0})^{2}}.
\ek
The resonance frequency of the above function takes place at
\bek
\dfrac{\text{d}}{\text{d}\omega}\Im \chi^{\text{LLG}}_{\pm}(\omega)=0\Rightarrow
\omega_{\text{res}}=\sqrt{1+\eta^{2}}\omega_{0},
\ek
while the FWHM amounts to
\bek
\Gamma= 2\eta\omega_{0}\label{eq:LLG-FWHM}
\dfrac{\sqrt{2+3\eta^{2}+2\sqrt{1+\eta^{2}}}}{1+\sqrt{1+\eta^{2}}}
\simeq 2\eta\omega_{0}.
\ek
We note that the above approximation is exact in the 
$\eta\rightarrow 0$ limit and involves only a $\sim10\%$ relative error for
$\eta=1$, which is by far the maximum value that damping can get for 
single adatoms;
for most of the elements analyzed in the main text we have 
$\eta\lesssim 0.5$~\cite{ibanez-azpiroz_zero-point_2016}, 
so the approximation of Eq. (\ref{eq:LLG-FWHM})
is indeed very good. Then, comparing Eq. (\ref{eq:LLG-FWHM}) to 
Eq. (\ref{eq:LLG-Ttrans}) we arrive to the relation between the FWHM and the transverse relaxation time
quoted in the main text:
\bek
T_{\perp}=\dfrac{2}{\Gamma}.
\ek

\section{Basic expressions of the Bloch-Redfield formalism}
\label{appendix:BR}

In this Appendix we provide a brief summary of the 
relaxation times in the context of Bloch-Redfield (BR) theory 
(see Ref.~\onlinecite{delgado_spin_2017} for 
details) in order to clarify the nomenclature regarding the 
relaxation time associated to an spin-excitation.
We begin with the longitudinal spin relaxation time, which in 
the BR theory describes the
decay rate of diagonal matrix elements of the reduced density operator and
is given by the following expression:
\bek\label{eq:appBRT1}
\dfrac{1}{T_{1}}\equiv \Gamma_{nm}=2\sum_{\alpha\beta}\Re\big(g_{\alpha\beta}(\omega_{mn})\big)
S_{\alpha}^{nm}S_{\beta}^{mn},
\ek
where $n,m$ label  the  electronic eigenstates of the adatom, $\alpha,\beta$ label the eigenstates of the substrate,
$\omega_{mn}=\epsilon_m - \epsilon_n$ with $\epsilon_i$ the eigenenergies,  
$g_{\alpha\beta}(\omega_{mn})$ is the substrate operator correlator and $S_{\alpha}^{nm}$
the matrix elements of the adatom's spin operator. 
Note that the term $\Gamma_{nm}$ in Eq. (\ref{eq:appBRT1}) corresponds to the scattering rate from state $n$ to $m$, hence
$T_{1}$ is associated to population transfer between different states. 
$T_{2}$, in turn, is termed as the decoherence time and describes the
decay rate of off-diagonal matrix elements of the reduced density operator. 
It can be separated into two different contributions,
namely the nonadiabatic one, $\gamma^{\text{nonad}}_{nm}$, and the adiabatic one, $\gamma^{\text{ad}}_{nm}$,
\bek\label{eq:appBRT2}
\dfrac{1}{T_{2}}=\gamma^{\text{nonad}}_{nm}+\gamma^{\text{ad}}_{nm}.
\ek
The nonadiabatic contribution is given by 
\bek\label{eq:app-nonad}
\gamma^{\text{nonad}}_{nm}=\dfrac{1}{2}\left(
\sum_{n'\neq n}\Gamma_{nn'}+\sum_{n'\neq m}\Gamma_{mn'}
\right),
\ek
while the adiabatic one reads
\bek\label{eq:app-ad}
\gamma^{\text{ad}}_{nm}=\dfrac{1}{2}\sum_{\alpha\beta}\Re\big(g_{\alpha\beta}(0)\big)
\Big( S_{\alpha}^{mm}-S_{\alpha}^{nn} \Big)\Big( S_{\beta}^{mm}-S_{\beta}^{nn} \Big).
\ek
Noteworthily, the relaxation time of an atomic spin-excitation in the BR theory 
is described by $\gamma^{\text{nonad}}_{nm}$ of Eq. (\ref{eq:app-nonad}). 
The most important aspect to note for our purpose here
is that $\gamma^{\text{nonad}}_{nm}$ in Eq. (\ref{eq:app-nonad})
involves a population transfer $\Gamma_{nn'}$, while 
$\gamma^{\text{ad}}_{nm}$ in Eq. (\ref{eq:app-ad}) does not. 
As a consequence, $\gamma^{\text{nonad}}_{nm}$ and hence the spin-excitation lifetime
is regarded as a $T_{1}$-like term 
(see Eqs. (69) and (70) of Ref.~\onlinecite{delgado_spin_2017}), 
even though it formally describes the decay rate of off-diagonal matrix elements of
the density operator rather than diagonal ones.
Meanwhile,  $\gamma^{\text{ad}}_{nm}$ is named  
the pure decoherence contribution~\cite{delgado_spin_2017}. 
This, in our understanding, is how and why 
the relaxation time of an atomic spin-excitation is 
associated to $T_{1}$ instead of $T_{2}$
in this context. 

We note that the above convention is not in line with the one 
adopted in the present work.
From our point of view, given that an atomic spin-excitation can be related to  
the damped precessional (transversal) motion of the adatom's magnetic moment, 
it is more natural to denote its lifetime by $T_{2}$ 
instead of with  $T_{1}$.

\section*{References}
\bibliography{biblio}

\end{document}